\documentclass[pdflatex, sn-mathphys-num]{sn-jnl}%

\usepackage{graphicx}%
\usepackage{multirow}%
\usepackage{amsmath,amssymb,amsfonts}%
\usepackage{amsthm}%
\usepackage{mathrsfs}%
\usepackage[title]{appendix}%
\usepackage{xcolor}%
\usepackage{textcomp}%
\usepackage{manyfoot}%
\usepackage{booktabs}%
\usepackage{listings}%
\usepackage[nohyperlinks, nolist ]{acronym}
\usepackage{braket}
\usepackage{tabularx}
\usepackage{import}
\usepackage{pgf}

\begin{document}

\begin{acronym}
\acro{qml}[QML]{Quantum Machine Learning}
\acro{ml}[ML]{Machine Learning}
\acro{tn}[TN]{Tensor Networks}
\acro{qtn}[QTN]{Quantum Tensor Networks}
\acro{ctn}[CTN]{Classical Tensor Network}
\acro{cm}[CM]{Confusion Matrix}
\acro{vqc}[VQC]{Variational Quantum Circuit}
\acro{mps}[MPS]{Matrix Product State}
\acro{tno}[TNO]{Tensor Network Operator}
\acro{mpo}[MPO]{Matrix Product Operator}
\acro{mpd}[MPD]{Matrix Product Disentangler}
\acro{nisq}[NISQ]{Noisy Intermediate Scale Quantum}
\end{acronym}

\title[Hybrid quantum tensor networks for aeroelastic applications]{Hybrid quantum tensor networks for aeroelastic applications}

\author*[1]{\fnm{M. Lautaro} \sur{Hickmann}}\email{lautaro.hickmann@dlr.de}

\author[1]{\fnm{Pedro} \sur{Alves}}\email{pedro.alves@dlr.de}

\author[2]{\fnm{David} \sur{Quero}}\email{david.queromartin@dlr.de}

\author[3]{\fnm{Friedhelm} \sur{Schwenker}}\email{friedhelm.schwenker@uni-ulm.de}

\author[1]{\fnm{Hans-Martin} \sur{Rieser}}\email{hans-martin.rieser@dlr.de}

\affil*[1]{\orgdiv{Institute for AI Safety and Security}, \orgname{German Aerospace Center (DLR)}, \orgaddress{\city{Ulm and St. Augustin},\country{ Germany}}}

\affil[2]{\orgdiv{Institute of Aeroelasticity}, \orgname{German Aerospace Center (DLR)}, \orgaddress{\city{G\"ottingen}, \postcode{37073}, \country{Germany}}}

\affil[3]{\orgdiv{Institute of Neural Information Processing}, \orgname{University of Ulm}, \orgaddress{\city{Ulm}, \postcode{89081}, \country{Germany}}}

\abstract{We investigate the application of hybrid quantum tensor networks to aeroelastic problems, harnessing the power of~\ac{qml}. By combining tensor networks with variational quantum circuits, we demonstrate the potential of~\ac{qml} to tackle complex time series classification and regression tasks. Our results showcase the ability of hybrid quantum tensor networks to achieve high accuracy in binary classification. Furthermore, we observe promising performance in regressing discrete variables. While hyperparameter selection remains a challenge, requiring careful optimisation to unlock the full potential of these models, this work contributes significantly to the development of~\ac{qml} for solving intricate problems in aeroelasticity. We present an end-to-end trainable hybrid algorithm. We first encode time series into tensor networks to then utilise trainable tensor networks for dimensionality reduction, and convert the resulting tensor to a quantum circuit in the encoding step. Then, a tensor network inspired trainable variational quantum circuit is applied to solve either a classification or a multivariate or univariate regression task in the aeroelasticity domain.}


\maketitle

\section{Introduction}\label{Intro}

Simulations of aeroelastic phenomena involve modelling complex fluid dynamics and the structural behaviour of components. For modern aircraft design, a detailed level of fidelity in the modelling of complex aeroelastic phenomena is essential. Increasing modelling fidelity leads to the need of resolving ever finer grids leading to an enormous computational effort for the numerical simulations. Therefore, finding efficient implementations is a key research field in aeroelastics. In particular, this involves developing techniques for the reduced order modelling of nonlinear aerodynamics. These techniques need to consider the complex nonlinear behaviour originated by the compressible, viscous and turbulent flow phenomena while not needing to simulate this behaviour on each grid point. Inherent difficulties are the nonlinear dependence and the high-dimensionality regarding the phase space on the grid required in order to describe such features.

Data driven implementations using machine-learning algorithms for aeroelastic simulations are currently under development as possible solutions to these requirements~\cite{NeuralNetworkCFD,NeuralNetworkBuffet}. Recently there has also been an increasing interest in utilising quantum computation and tensor network approaches (both on classical and quantum hardware) for~\ac{ml}~\cite{EdwinStoudenmire, reyes2020multiscaletensornetworkarchitecture, Dilip2022,shen2024classificationfashionmnistdatasetquantum, Huggins.2019}. Therefore, we investigated the prospect of using hybrid quantum tensor network based algorithms for aeroelastic problems. 

A wide variety of \ac{qml} approaches employing quantum circuits with tunable parameterised gates, so called~\acp{vqc}, have recently been proposed~\cite{Schuld}. Quantum tensor network for ML can be realised by~\acp{vqc} using a tensor network inspired internal gate structure. \ac{tn} were initially developed to reduce the computational cost of lowly entangled multi-particle quantum states. Nevertheless, they are able to efficiently approximate a wide variety of large tensorial objects using a regular, less complex structure. Thus, providing a convenient approach to \ac{qml}~\cite{Rieser2023}. 

Our work focuses on advancing hybrid~\ac{qml} methods, with a particular emphasis on efficient data encoding into quantum circuits. Efficient encoding is a crucial step in QML. Recently, two promising approaches have emerged: pre-training quantum circuits to approximately encode data~\cite{shen2024classificationfashionmnistdatasetquantum}, and using~\acp{tn} based encoding techniques to exactly encode data into quantum circuits~\cite{PhysRevA.101.032310}. However, these techniques have previously been used as preprocessing steps, where data is encoded once and then used as input for trainable quantum circuits. In contrast, our approach integrates the~\ac{tn}-based encoding into a fully end-to-end trainable hybrid algorithm. This approach entails a~\ac{tn} decomposition of the classical step to quantum gates, as explained in~\autoref{sec:TN_pre}. 

Our setup enhances current~\ac{qml} algorithms by combining three key components: a trainable~\ac{tn}-based dimensionality reduction,~\ac{tn}-based data encoding, and a trainable~\ac{tn}-inspired~\ac{vqc}~\cite{shen2024classificationfashionmnistdatasetquantum}, as explained in~\autoref{sec:Q-classifier}. This integrated approach enables end-to-end training using a single classical optimiser, allowing us to solve regression and classification tasks. An additional major technical contribution is the inclusion into a thorough state-of-the-art machine learning pipeline, including optimisation tools, mini-batch training and hyperparameter search with cross-validation.

\section{Methods}\label{Met}

\subsection{Aeroelastic Application}\label{sec:ApplicationCase}

Classical tensor networks have various application scenarios within aeroelastics~\cite{VolterraTensorNetwork}. Relevant applications  include data-driven aeroservoelasticity or the computation of dynamic loads resulting in an airframe of a manoeuvring aircraft~\cite{DataDrivenAFS}. 
Within this field, nonlinear effects originating either from the structural or aerodynamic counterparts or a combination of them are of importance. The overall goal is to derive models from data which are able to predict aeroelastic characteristics (including damping) and thus, the stability behaviour of the system~\cite{OperationalModalAnalysis}.

 One problem of particular interest in aeroelasticity is the determination of the flutter stability. To determine the stability, the feedback interaction between the structure and the aerodynamic forces has to be considered including inertial and elastic forces. We use a simplified aeroelastic configuration including a low-dimensional aerodynamic model for investigating the potential of~\ac{qml} for estimating the flutter stability of the system, based on~\cite{ASELoewner}. 

The selected case comprises a typical aeroelastic section of a wing with three degrees of freedom including heave $h$ (positive downwards), pitch $\theta$ (positive nose up) around the elastic axis location and an aileron rotation $\beta$ (positive with trailing edge down)~\cite{Tewari2015}. No assumption regarding the flow physics has been made and thus the methods are entirely data-driven, similar to~\cite{DataDrivenAeroelasticityTurbomachinery}. The corresponding aeroelastic equations of motion are given by

\begin{align}
\left[\begin{array}{ccc}
1 & x_{\theta} & x_{\beta}\\
x_{\theta} & r_{\theta}^{2} & r_{\beta}^{2}+x_{\beta}\left(c-a\right)\\
x_{\beta} & r_{\beta}^{2}+x_{\beta}\left(c-a\right) & r_{\beta}^{2}
\end{array}\right]\left[\begin{array}{c}
\left(\frac{1}{L_\text{ref}}\right)\frac{d^{2}h}{dt^{2}}\\
\frac{d^{2}\theta}{dt^{2}}\\
\frac{d^{2}\beta}{dt^{2}}
\end{array}\right]&+\left[\begin{array}{ccc}
\omega_{h}^{2} & 0 & 0\\
0 & r_{\theta}^{2}\omega_{\theta}^{2} & 0\\
0 & 0 & r_{\beta}^{2}\omega_{\beta}^{2}
\end{array}\right]\left[\begin{array}{c}
\frac{h}{L_\text{ref}}\\
\theta\\
\beta
\end{array}\right]\label{eq:AeroelasticEquation}\\
&=\frac{1}{\pi\mu}\left(\frac{U_{\infty}}{L_\text{ref}}\right)^{2}\left[\begin{array}{c}
-c_{l}\left(t\right)\\
2c_{m}\left(t\right)\\
2c_{\beta}\left(t\right)
\end{array}\right],\nonumber
\end{align}

\noindent where the structural damping has been neglected. The aerodynamic forces acting upon the structure are represented by the lift coefficient $c_{l}$, the pitching moment at the elastic axis $c_{m}$ and the hinge moment at the aileron hinge axis $c_{\beta}$. A set of parameters has been chosen to be constant and their values are specified in~\autoref{tab:ConstantParameters}, where the non-dimensional distances are obtained upon dividing by the reference length $L_\text{ref}$. \autoref{tab:VariedParameters} shows the variation of parameters carried out for the applications described next.

\begin{table}[ht]
\begin{centering}
\begin{tabular}{ccc}
\hline 
Description & Parameter & Value\\
\hline 
\hline 
Reference length & $L_\text{ref}$ & $0.5$ (m)\\
Uncoupled heave natural frequency & $\omega_{h}$ & $50$ (rad/s)\\
Uncoupled pitch natural frequency & $\omega_{\theta}$ & $100$ (rad/s)\\
Uncoupled aileron natural frequency & $\omega_{\beta}$ & $300$ (rad/s)\\
Non-dimensional distance from e.a. to the airfoil c.g. & $x_{\theta}$ & $0.2$\\
Non-dimensional distance aileron h.a. to aileron c.g. & $x_{\beta}$ & $0.0125$\\
Non-dimensional airfoil radius of gyration about e.a. & $r_{\theta}$ & $\sqrt{0.25}$\\
Non-dimensional aileron radius of gyration about aileron h.a. & $r_{\beta}$ & $\sqrt{0.00625}$\\
Non-dimensional distance between the midchord and the aileron h.a.  & $c$ & $0.5$\\
\hline 
\end{tabular}
\end{centering}
\caption{Constant parameters. Centre of gravity is denoted by c.g., elastic axis by e.a., and hinge axis by h.a.\label{tab:ConstantParameters}}
\end{table}

\begin{table}[ht]
\begin{centering}
\begin{tabular}{ccc}
\hline 
Parameter & Interval & Increment\tabularnewline
\hline 
\hline 
Non-dimensional distance between midchord and e.a. & $a\in\left[-0.4,0.4\right]$ & $\bigtriangleup a=0.1$\tabularnewline
Mass ratio & $\mu\in\left[10,50\right]$ & $\bigtriangleup\mu=0.1$\tabularnewline
Airspeed & $U_{\infty}\in\left[150,350\right]$ (m/s) & $\bigtriangleup U_{\infty}=1$ (m/s)\tabularnewline
\hline 
\end{tabular}
\par\end{centering}

\caption{Varied parameters. Elastic axis is denoted by e.a.\label{tab:VariedParameters}}
\end{table}

\autoref{eq:AeroelasticEquation} cannot be directly numerically integrated in time, as the aerodynamic coefficients are provided in the frequency-domain. When considering incompressible two-dimensional unsteady potential flow, they are provided in the frequency domain as irrational functions of the frequency. Thus, a specific procedure is applied in order to transform it into a state-space representation \cite{ASELoewner}, which can finally be numerically integrated in time with a common ordinary-differential equation (ODE) solver:
\begin{align}
\frac{d}{dt}\left(\left[\begin{array}{c}
\mathbf{u}_{h}\\
\frac{d\mathbf{u}_{h}}{dt}\\
\mathbf{x}_{a}
\end{array}\right]\right)&=\mathbf{A}_{ae}\left[\begin{array}{c}
\mathbf{u}_{h}\\
\frac{d\mathbf{u}_{h}}{dt}\\
\mathbf{x}_{a}
\end{array}\right],\label{eq:StateSpaceAE}\\
 \mathbf{u}_{h}&=\mathbf{C}_{ae}\left[\begin{array}{c}
\mathbf{u}_{h}\\
\frac{d\mathbf{u}_{h}}{dt}\\
\mathbf{x}_{a}
\end{array}\right],\nonumber 
\end{align} 

\noindent where $\mathbf{u}_{h}=\left[h/L_\text{ref}\,\,\theta\,\,\beta\right]^{T}$ and $\mathbf{x}_{a}$ contains the resulting aerodynamic states. For details on the matrices $\mathbf{A}_{ae}$ and $\mathbf{C}_{ae}$ the interested reader is referred to \cite{ASELoewner}. Once written in this form, the eigenvalues of the matrix $\mathbf{A}_{ae}$ determine the flutter stability of the aeroelastic system, which is then dependent on the value of the parameters given in~\autoref{tab:VariedParameters}, provided the parameters in~\autoref{tab:ConstantParameters} haven been fixed.

\begin{figure}[t]
    \centering
    \resizebox{119mm}{!}{\input{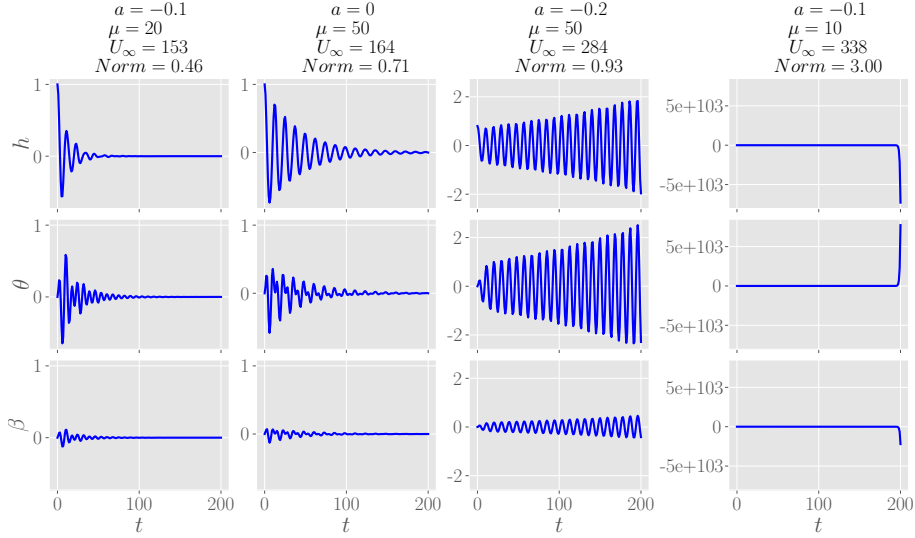}} 
    \caption{Prototypical aeroelastic time series responses for each degree of freedom for four sets of aeroelastic parameters. The two on the left are stable and the two on the right unstable responses.} \label{fig:curves}
\end{figure}

The goal of this application case is to apply hybrid quantum algorithms to the tasks of stability classification based on time series and regression of parameters from a time series. The stability of the system described in~\autoref{eq:AeroelasticEquation} is considered when subjected to non-zero initial conditions. In particular, the initial value of the first state component corresponding to a heave displacement is set to $1$. Note that the physical magnitude is not of relevance here, as~\autoref{eq:StateSpaceAE} is linear with respect to the states. Four representative time histories are provided in~\autoref{fig:curves}, where two stable and unstable cases each are represented for different combinations of the parameters $\left(a,\mu,U_{\infty}\right)$, the three time series describe the response on each degrees of freedom $(h, \theta, \beta)$.

\subsection{TN based preprocessing and encoding}\label{sec:TN_pre}

\acp{ctn} have various applications within aeroelastics, such as aeroelastic system identification. The goal is to derive data driven models which enable the prediction of aeroelastic characteristics including the stability behaviour of the system~\cite{OperationalModalAnalysis}.

\begin{figure}[t]
    \centering
    \resizebox{119mm}{!}{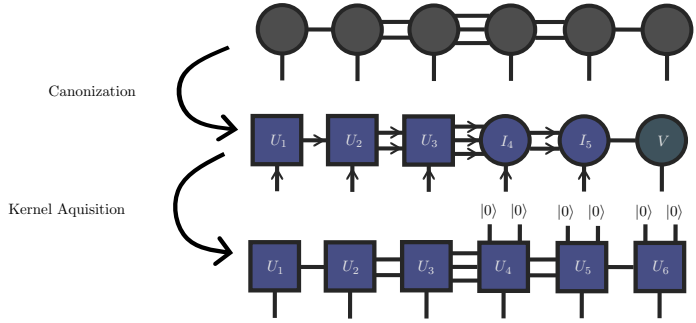}
    \caption{\ac{qtn} quantum circuit mapping example of a \ac{mps} representing a quantum state where the bond dimension increases exponentially to the centre bond. Here, the main phases are highlighted: the canonisation already leaves 3 tensors as unitary gates (blue squares), 2 tensors as isometries (blue circles) and 1 orthogonality centre representing essentially a normalised vector (green circle); the kernel acquisition step ensures that the full unitary gates can be found and the beginning of the quantum wires can be assigned with the zero-state $ \ket{0} $ .}~\label{fig:natrual_mps_to_circuit}
\end{figure}

Tensor Networks were originally conceptualised to facilitate the simulation of Quantum Many-Body Systems by reducing the amount of correlations inside a quantum state~\cite{wall2022tensor}. They are a set of tensor objects connected with each other in a specific layout through index contraction. Several different regular tensor network layouts with varying dimensionality have been studied. The most common ansatz is the~\ac{mps}~\cite{MPS} shown in the first row of~\autoref{fig:natrual_mps_to_circuit}. Other layouts are tree tensor networks (TTN)~\cite{TTN}, MERA networks~\cite{MERA}, which are trees with entangled branches and two-dimensional PEPS networks~\cite{PEPS}.

The aforementioned~\acp{tn}, are a powerful tool for representing complex data structures, enabling the efficient manipulation of classical and quantum systems. By applying~\acp{tno}, it is possible to perform operations on data in~\ac{tn} format, effectively modifying the underlying structure. \acp{tno} represent local linear transformations, akin to matrix multiplications, which can be applied to specific sections of the~\ac{tn}. For instance,~\acp{mps} can be transformed using~\acp{mpo}, which are defined by introducing an additional free index at each site, where one is considered the upper index (free input index) and lower index (free output index). By contracting the input indices of the~\ac{mpo} with those of the~\ac{mps}, a new, transformed~\ac{mps} is produced from the free output indices, allowing for efficient manipulation and analysis of complex data structures.

When choosing a tensor network type for a specific task, it is crucial to take the scaling behaviour of the task into account. While one-dimensional data like time series are handled well using MPS, images require a two-dimensional scaling of the information entropy in the worst case. In this study, we focus on time series therefore a one-dimensional MPS layout is well suited. In an MPS two additional hyperparameters can be chosen: the bond dimension which is determined by the number of qubits passed on from each node to the next, and the number of data qubits that are passed to the circuit at each iteration.

A \ac{qtn} is a \ac{tn} that represents a compressed version of a quantum state implying that the \ac{tn} is primarily under the normalisation condition. One additional property relevant for this work is that a canonically gauged one-dimensional \ac{qtn} can be mapped to a quantum circuit \cite{PhysRevA.101.032310, liu2019machine, huggins2019towards}. Once in canonical gauge, the majority of its tensors are isometries which can be converted into unitary gates with some linear algebra kernel acquisition technique \cite{PhysRevA.101.032310}. In other words, the output of such circuit of unitaries reflecting the one-dimensional layout, is the quantum state encoded by the \ac{qtn}. \autoref{fig:natrual_mps_to_circuit}, shows the mapping of a canonised \ac{mps}, to a quantum circuit.

\acp{tn} already found their way into applications within classical machine learning \cite{reyes2020multiscaletensornetworkarchitecture}. For instance, here a \ac{ctn} can represent an input vector, a linear operator or encode non-linear functions while benefiting from local operators that preserve a compressed representation of the problem at hand. At the expense of a normalisation constant, a \ac{ctn} can be transformed into a \ac{qtn}, for it to subsequently be mapped to a quantum circuit.

For our use case, the input data consists of a three-dimensional time series described in~\autoref{sec:ApplicationCase}. The time series have a wide range of values with the converging ones usually being in the range of $\pm 1$ but the diverging one can take values over $\pm 5 \times 10^{100}$. As a first step we normalise each group of time series as one, i.e. concatenate all three time series, normalise the resulting one and then separate them again. This way the amplitude relation between each dimension of the time series remains unchanged. Each data point, i.e. each group of three time series for a set of aeroelastic parameters $(a, \mu,U_\infty)$, is normalised independently and the norms (re-scaled to $[0, \pi]$) are saved to be utilised as an input to the quantum circuit.

\begin{figure}[t]
    \centering
    \resizebox{119mm}{!}{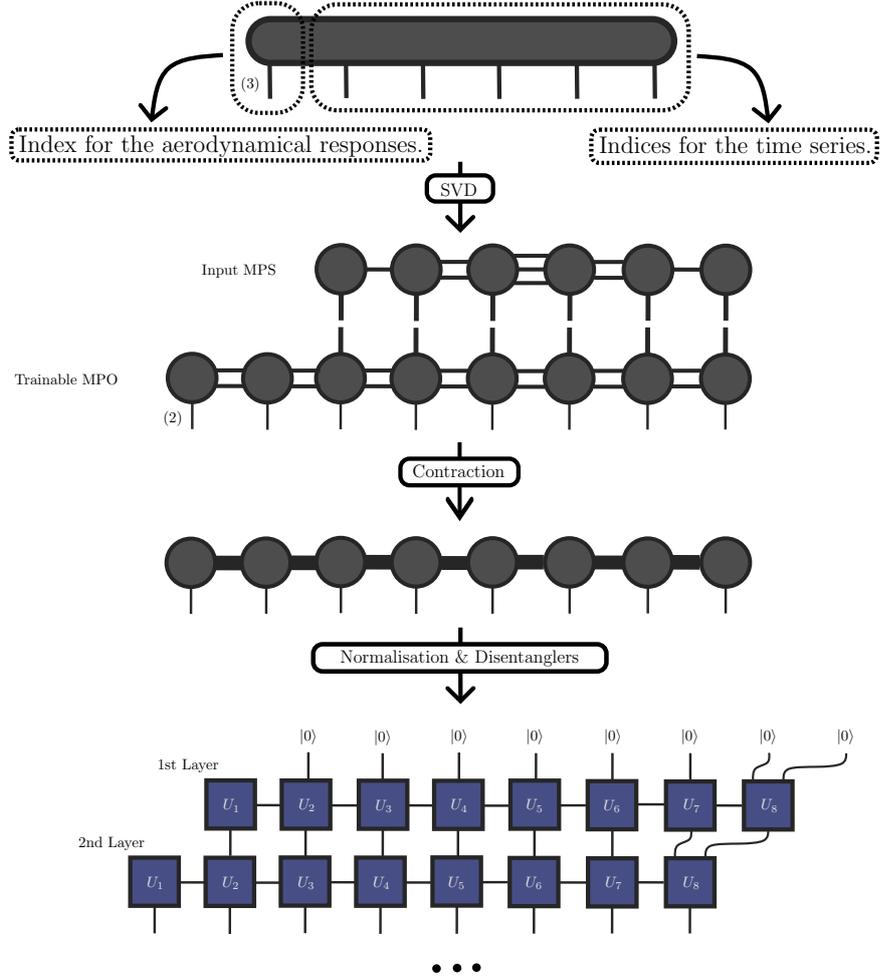}
    \caption{Preprocessing and encoding structure. First we encode the three time series comprising each data point into an~\acs{mps}. We then apply an~\acs{mpo} based dimensionality reduction and non-linearities. Finally we disentangle the resulting~\acs{tn} into a quantum circuit.}~\label{fig:HTN}
\end{figure}

As quantum mechanics is a linear theory, nonlinear behaviour can only be introduced by interactions with the classical environment, e.g. by carrying out measurements or during encoding of the normalised inputs $x$ using some encoding map $\Phi(x)$~\cite{Yan2015}. Finding the right encoding strategy for a \ac{qml} application is often a difficult task. A variety of methods have been developed which lie between the two extremes: ``qubit efficient'', realised by amplitude encoding, and ``gate efficient'', realised by binary encoding. In this work, we decided to use a tensor network operator to reduce the dimensionality of the data and learn the encoding to be deployed on the quantum circuit, as shown in~\autoref{fig:HTN}. Additionally, tensor networks mapped to quantum circuits can have qubit efficient representations~\cite{Rieser2023}.

Due to the one-dimensional structure of the time series, tensor networks and specifically~\ac{mps} are well suited to express this type of data~\cite{Rieser2023}. To improve the compatibility of our data with~\acp{tn}, we preprocessed the original time series by upsampling it from $201$ time steps to $3^5=243$ using smoothing B-splines. This upsampling enables an efficient decomposition into a $5$-node~\ac{mps} with free indices of dimension $3$, allowing for a compact representation of the data. An additional free index is introduced to select one of the three time series, effectively encoding the time series dimension. The preprocessing steps are illustrated in the first row of~\autoref{fig:HTN}.

This~\ac{mps} is contracted with a subsequent trainable~\ac{mpo} that aims to reduce the dimensionality of the input~\ac{mps} and learns the optimal encoding scheme to the quantum circuit. For this purpose, the \ac{mpo} has $ 6 $ free input indices of dimension $ 3 $ and $ 8 $ free output indices of dimension $ 2 $. The dimension of the \ac{mpo}'s internal bonds can be adjusted to carry more trainable parameters and potentially add more correlations to the data, thus being considered a hyper-parameter. Naturally, the output of this contraction is another~\ac{mps} with $ 8 $ free indices of dimension $ 2 $. To improve the expressivity, we applied a \textit{tanh} nonlinearity on the resulting~\ac{mps} parameters before normalisation. 

Since one of the objectives of this preprocessing step is to learn what to encode in the quantum circuit, this output \ac{ctn} still needs to be converted into a \ac{qtn} which can be done by normalising it~\cite{Dborin2022}. We then used the technique shown in \cite{PhysRevA.101.032310}, to map it into a quantum circuit. This technique adds layers of cascading $2$-qubit gates that have the ability to progressively disentangle the state represented by the output of the~\ac{mpo} (except for a normalisation constant) to the zero-state $ \ket{00...} $, hence being called \ac{mpd}. From another perspective, inverting the order of these layers and transpose-conjugating every unitary, the layers will progressively entangle the state $ \ket{00...} $ until the desired state is reached. The more layers there are, the more correlations/entanglement can be achieved in the state. Therefore, the number of~\ac{mpd} layers is considered a hyper-parameter in this set-up for gradually adjusting the entanglement of the input state to the~\ac{vqc} circuit. The complete procedure is shown in~\autoref{fig:HTN}.

\subsection{TN inspired VQC classifier}\label{sec:Q-classifier}

Once we have the data encoded into a quantum circuit, the next step is to process it using~\ac{qml}. Quantum computation in general and~\ac{qml} in particular are still in very early stages of development. Today, most quantum algorithms are written on a circuit level. When designing a quantum circuit, choices on a very basic level must be made, e.g. the data encoding circuit, entangling schemes and the measurement processes. As it is not clear to date which choices are most relevant for the quantum machine learning application we carried out a comprehensive hyperparameter search.

 A~\ac{vqc} is a quantum circuit with gates that feature tunable parameters, usually rotational gates. A general variational unitary can be decomposed into a combination of rotation and entangling gates like the CNOT. A common category of~\ac{vqc} architectures for machine learning are layered~\acp{vqc}. Here, the circuit consists of encoding blocks that map the data to the circuit, and variational blocks which entangle the qubits and introduce the optimisable parameters. To increase expressivity, these blocks are executed repeatedly~\cite{Schuld}.

\begin{figure}[t]
\centering
\resizebox{119mm}{!}{\input{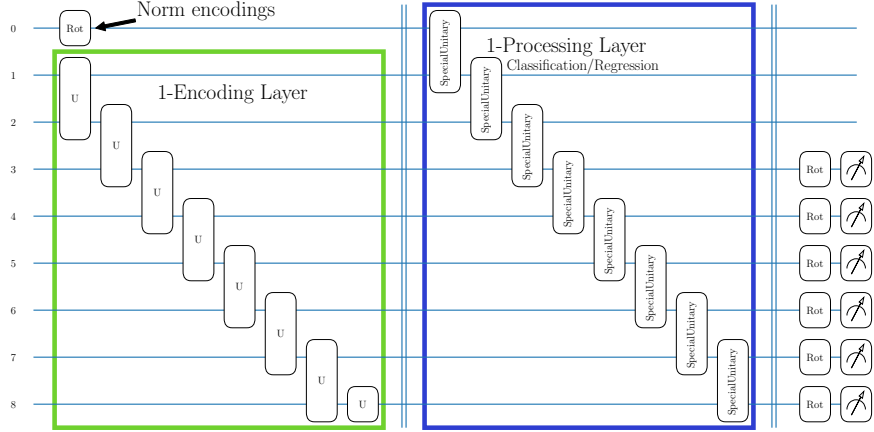}} 
\caption{Typical layout for an~\ac{mps} inspired \ac{vqc} for the regression task measuring multiple qubits. Here including Norm encoding and using one encoding and one classification layer. Lastly a trainable measurement layer is applied. Similar to~\cite{Fischbach2025}.}~\label{fig:TNiVQC}
\end{figure}

Another approach is to employ a~\ac{qtn} for machine learning using parameterised gates, often also called~\ac{tn} inspired ansatz. It is a variety of~\ac{vqc} that carries a tensor network based internal gate structure~\cite{Dilip2022}. The construction of the tensor network approach using ``states'' and ``operators'' makes it straightforward to translate the concept to quantum computation as seen in~\autoref{fig:TNiVQC}. Both are realised by a set of parameterised multi-qubit gates where the only difference between states and operators is whether the gates have only incoming or outgoing free bonds or both.

To make use of the capabilities of a quantum computer, the individual qubits have to be entangled by using multi-qubit gates. These gates have several free parameters that can be used to define how the incoming data is processed. When kept trainable, they can act as parameters of a machine learning algorithm. 

As shown in~\autoref{fig:TNiVQC}, we first encoded the data by using one or more layers of arbitrary unitaries derived from the preprocessing~\ac{mps} through the disentangling process. Additionally, we encoded the norm of the original time series obtained in the first normalisation step and the norm of the resulting~\ac{mps} as parameters of a general rotation on the first qubit. Since otherwise, the information on the amplitude relationship between time series would have been lost. 

After the encoding, we applied a~\ac{mps} inspired structure constructed from iterated layers of two qubit gates~\cite{Dilip2022,Jobst2024efficientmps,shen2024classificationfashionmnistdatasetquantum}. For the two qubit gates, we considered strongly entangling layers on two qubits~\cite{PhysRevA.101.032308} with $6$ trainable parameters per gate and general $\mathrm{SU(4)}$ unitaries~\cite{Wiersema2024herecomessun} with $15$ trainable parameters per gate. For~\ac{qtn}, some authors recommend shuffling the remaining virtual qubits before measurement using a single layer of general rotations on the qubits being measured~\cite{shen2024classificationfashionmnistdatasetquantum}. 

The measurement which later can be interpreted as the result of our machine learning algorithm can be performed in several ways. For classification problems, the simplest way is to chose one qubit as the output qubit. We measured the probabilities of the basis states on the last qubit and interpreted them as class probabilities, which were then compared against one hot encoded binary stability labels. 

For the regression task, we measured either the expectation values of individual qubits or tensor product of pairs of qubits to predict the aeroelastic parameters. To make the comparison possible, we re-scaled the aeroelastic parameters to the $\pm 1$ range independently per feature. We investigate univariate regression by predicting each of the aeroelastic parameters with different models and multivariate regression by predicting all aeroelastic parameters at once. For the latter case, we measured either three $Z_i$ observables or three tensor products $\{Z_i \otimes Z_j\}_{i\neq j}$ observables on non-overlapping pairs of qubits and interpreted them as elements of the target vector. Similarly for the univariate setup, we only measured one observable.

Besides the architecture of the quantum circuit, other parameters can be adjusted. There are several different classical optimisers that can be uses to train hybrid-quantum circuits. The parameters of the~\ac{mpo} and~\ac{vqc} where then jointly optimised using a classical optimiser through backpropagation with auto-differentiation. We used the well known Adam optimiser~\cite{kingma2017adam}, which uses global gradients. As loss functions, we used cross entropy for the classification tasks, and the Huber loss~\cite{huber_loss} for the regression tasks. 

The computations were performed using the Quimb~\cite{gray2018quimb} library for~\acp{tn}, PennyLane~\cite{bergholm2022pennylaneautomaticdifferentiationhybrid} for quantum circuits and simulations, and JaX~\cite{jax2018github} for the~\ac{ml} components. For statistical robustness we used cross validation, using the ShuffleSplit method. We used the implementations provided in the scikit-learn library~\cite{scikit-learn}. Lastly we used the Optuna~\cite{optuna} hyperparameter search framework.

For each hyperparameter configuration, a $5$-fold cross validation was carried out, using the methods previously explained. At the end of each training, the maximum scores per fold over all previous epochs of the metrics were averaged and used as the objective value for the hyperparameter search algorithm. We used the well known $F_1$ score for the classification task and $R^2$ score for the regression task.

After conducting the hyperparameter search, we retrained the best configuration for each task using the complete training set as folds, i.e. we ran the training $5$ times with different random seeds on the complete training dataset (training + evaluation datasets used for the hyperparameter search), and tested the trained models on the test set.

\section{Results and Discussion}\label{sec:Res}

\begin{figure}[t]
\centering
\resizebox{119mm}{!}{\input{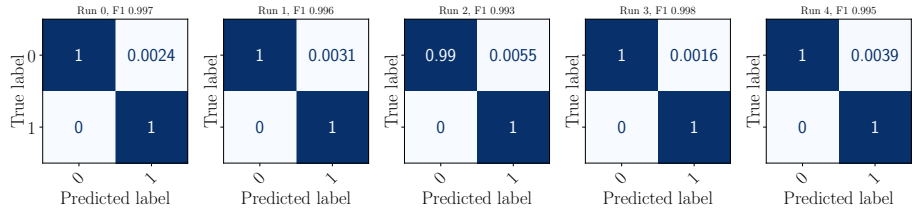}} 
\caption{F1 confusion matrices on the test set for the best model for the classification task, for $5$ different random training seeds.}~\label{fig:CMs}
\end{figure}

\begin{figure}[b]
\centering
\resizebox{119mm}{!}{\input{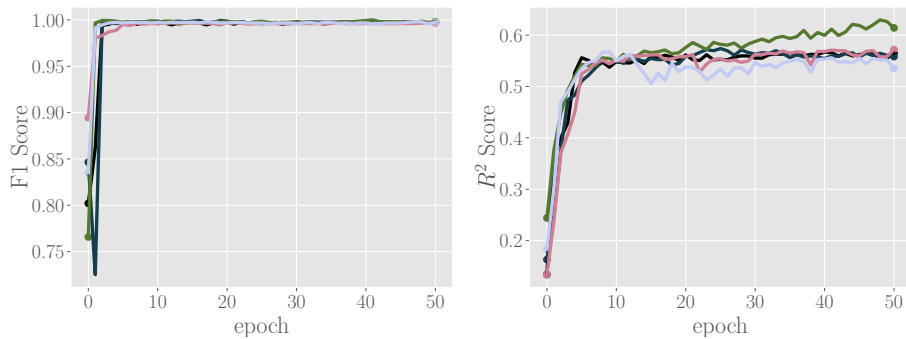}} 
\caption{Test F1 score (left) and test $R^2$ score (right) values over training epochs for $5$ different training seed of the best classification (left) and multivariate regression (right) model.}~\label{fig:TCs}
\end{figure}

We found that our hybrid~\ac{tn} inspired algorithm could easily solve the binary time-series classification, achieving a maximum $F_1$-score of well above $0.9$, averaged over $5$ repeated training runs. The best model achieved a $F_1$-score of $0.998$, as shown in the Confusion Matrices (CMs) in~\autoref{fig:CMs}. The models generalised very well, and we observed no overfitting. We carried out a very limited hyperparameter search, since we quickly found well performing configurations. As it can be observed in the~\acsp{cm}, all training seeds converged towards good results.

While doing the hyperparameter search, we could observe that many configurations were unstable, achieving significantly different results for each fold. Overall, several configuration achieved good results, the best model used only a small~\ac{mpo} bond dimension of $2$ and only one disentangling layer, but needed four~\ac{tn} inspired quantum classification layers. This hints at the majority of the processing being done on the~\ac{vqc} side, for a highly compressed and potential low-entanglement representation obtained through the utilised~\ac{mpo}.

Our analysis of the training behaviour revealed that most runs converged rapidly, with most models achieving optimal performance within $5$ epochs, with the amount of gradients updates depending on the utilised batch size. Although this might suggest the presence of barren plateaus, a closer examination of the gradient variance showed that the quantum part of the model exhibited a tendency towards zero, but did not completely disappear. In contrast, the gradients of the~\ac{mpo} displayed more instability, with some values converging towards zero before spiking up and then returning to near zero. Despite this unusual behaviour, the classifier achieved outstanding results. This behaviour calls for an in depth analysis of the loss landscape, and in particular the relations between the hyperparameters and the trainability of the models.  

The test behaviour over training epochs for the best hyperparameter configuration on the classification task, as shown on the left side of~\autoref{fig:TCs}, demonstrate rapid convergence, with approximately $2$ epochs required to achieve stable performance. This corresponds to around $114$ mini-batch gradient updates, for the selected batch size of 128. Notably, the models' performance did not degrade over prolonged training. Furthermore, the initial spread of F1 values at epoch $0$, reflects the impact of different weight initialisation. However, as training progressed, all models converged to remarkable results, suggesting that the optimisation process was robust and effective.

\begin{figure}[t]
        \centering
        \resizebox{119mm}{!}{\input{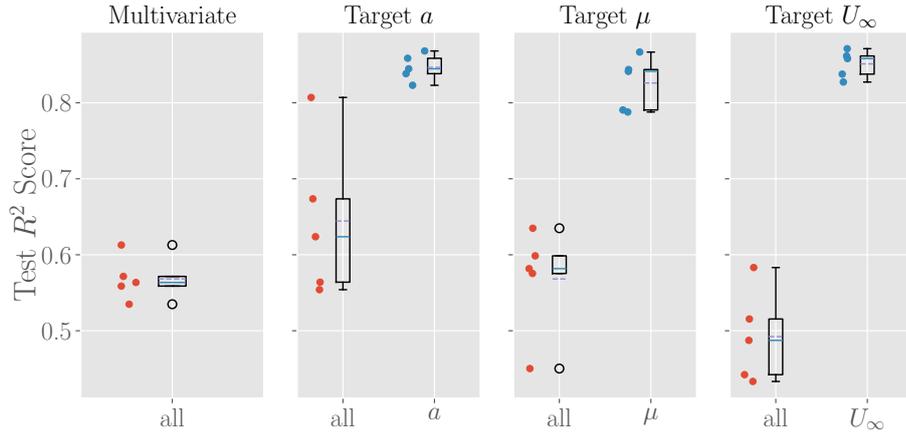}} 
		\caption{Test regression score for the best model for the multivariate and univariate regression task, averaged over $5$ runs of each model.
		Each column represents different prediction targets, and the labels at the bottom represent the targets the model was trained to predict.}\label{fig:reg}
\end{figure}

For the regression task, we used the same~\ac{tn} inspired~\ac{vqc} algorithm as for classification, with adjustments for the number of measured qubits and the type of measurement carried out. As described in~\autoref{sec:Q-classifier}, the main difference is that we measured expectation values instead of basis state probabilities. Since we require more granular control over the predictions. 

\begin{figure}[t]
        \centering
        \resizebox{119mm}{!}{\input{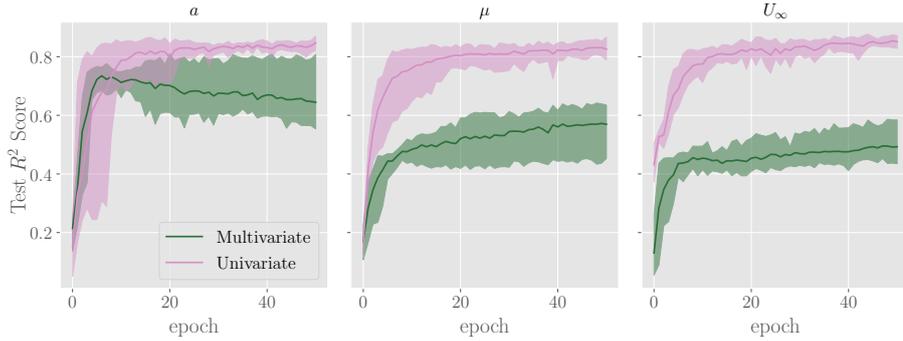}} 
		\caption{Test regression score for the best model for the multivariate and univariate regression task over training epochs, averaged over $5$ runs of each model.
		Each column represents different prediction targets.}\label{fig:reg_curves}
\end{figure}

Overall, the regression task exhibited less stability, particularly for multivariate regression, as shown in the right side of~\autoref{fig:TCs}. The univariate regression results were significantly better, as depicted in~\autoref{fig:reg} and~\autoref{fig:reg_curves}. 

A notable observation from the hyperparameter search was that some multivariate configurations achieved near-optimal performance on specific target parameters (one of $a, \mu$ or $U_\infty$), comparable to those of univariate models. However, these models often predicted a single target dimension well while failing to accurately predict others. Nevertheless, the best multivariate model achieved a more balanced performance, albeit with a slightly lower $R^2$ score per target dimension. This suggests that while the model was able to capture only certain aspects of the data, but not everything necessary for performing all three regressions at once.

In contrast, the models trained in a univariate fashion, i.e. to predict only a single target dimension, achieved outstanding results. Notably, different hyperparameter configurations yielded the best results for different target dimensions, suggesting that the chosen ansatz lacked flexibility. This may be attributed to the lack of trainable non-linearities.

As observed in~\autoref{fig:reg} and~\autoref{fig:reg_curves}, the repeated runs for different seeds exhibited good convergence, but with a considerable spread. This variability implies that better initialisation techniques are needed. Our results indicate that using a noisy identity initialisation led to more stable results compared to a random uniformly distributed initialisation. However, a more in-depth analysis is required to isolate the factors contributing to this behaviour.

An interesting observation was made when examining the test $R^2$ scores during training of the best selected model, shown in~\autoref{fig:reg_curves}. The multivariate model initially performed well on the $a$-dimension, but then its performance degraded. Conversely, the other two dimensions ($\mu$ and $U_{\infty}$) showed improvement, suggesting that the model was learning to balance its predictions. This phenomenon may indicate that the model was initially over-specialised to one dimension and then generalised to the others. This behaviour could indicate at a lack of expressibility of the model, or the hardness of the multivariate regression task.

The multivariate models exhibited a significantly wider spread across the $5$ different training seeds compared to the univariate setup. Notably, the two setups displayed opposing trends in variance over time. The multivariate model showed relative stability at epoch $10$, followed by a significant divergence, particularly for the $a$ and $\mu$ dimensions. In contrast, the univariate models demonstrated a reduction in variance over advanced training epochs.

The difference on the $R^2$ scores for multivariate regression and the reason the hyperparameter configurations differ for each univariate setup can be explained by the different granularity of each component: $a$ has $9$ distinct values, $\mu$ has $5$ and $U_\infty$ has $201$. The score difference between $a$ and $\mu$ could be explained by the former having enough values to better cover the normed target value range, resulting in smaller errors between predictions and targets.

The regression task proved to be considerably more challenging than the classification task. The prediction of discrete variables with multiple possible values using expectation values of observables is a more complex task than binary classification. The model struggled to accurately predict the values, and the random initialisation of the models resulted in significant variability in performance.

In particular, we observed that for the cross validation some runs of the same configuration worked out well while others did not converge to a good score. This behaviour could be explained by instabilities arising from the data points randomly selected for each cross-validation fold, i.e. the failure to generalise from specific subsets of the dataset. This may be mitigated by using more than five runs, i.e. using more folds, per configuration. Although, steeply increasing the resources requirements accordingly. The second possibility would be to theoretically investigate the loss landscapes and initialisation of such quantum circuits and use these results to begin with better initial conditions.

The analysis of hyperparameter importance revealed similarities between the classification and regression tasks, as well as between the two regression setups. Overall, the interpretation of the hyperparameter search results was challenging due to the high spread for single hyperparameter values. This lead to the conclusion that not single, but combinations of hyperparameters have a significant impact on the model's performance.

We used four parameter importance estimators, namely fANOVA~\cite{pmlr-v32-hutter14}, Shapley TreeExplainer~\cite{lundberg2020local2global}, Mean Decrease Impurity (MDI)~\cite{agarwal2023mdi}, and PED-ANOVA~\cite{watanabe2023pedanovaefficientlyquantifyinghyperparameter} to select a subset of most important hyperparameters. The results showed that in almost all cases the number of quantum processing layers, the learning rate, and the chosen quantum ansatz were the most significant contributors to the model's performance, except for the univariate $a$ model were instead of quantum layers the~\ac{mpo} bond dimension played a role. 

These observations suggest that the~\acp{vqc} play a crucial role in determining the model's performance. Two possible interpretations are that either the~\acp{vqc} are doing most of the processing, and their performance is heavily influenced by these hyperparameters. Or, alternatively, the~\acp{vqc} may be a bottleneck towards the model's information processing capacity. Especially for the more complex regression tasks since the classification tasks could be perfectly solved. 

On the other hand, the results also suggest that the~\ac{mpo} parameters have a limited impact on the model's performance. This could either mean that all tested configurations discard too much data leading to the quantum part having to do the heavy lifting, or that they provide enough flexibility to adapt to the needed information content. 

The analysis highlights the importance of hyperparameter tuning and the need for further investigation: An in-depth ablation study would be necessary to fully understand the impact of these hyperparameters on the model's performance. Finding out what would be the best way to improve the synergies of the classical,~\acp{tn}, hyperparameters with the quantum circuit hyperparameters remains challenging, and an in depth information theoretical analysis is needed. 

Overall, we show that realistic regression problems can be tackled by hybrid QML approaches and the results provide valuable insights into the importance of hyperparameter tuning and the challenges of optimising~\acp{vqc}. Further research is needed to fully understand the impact of these hyperparameters on the model's performance.  

\section{Conclusion and future work}\label{Con}

In conclusion, our study demonstrates the successful application of hybrid quantum tensor network-based algorithms to aeroelastic problems. By integrating three key components - trainable~\ac{tn}-based dimensionality reduction,~\ac{tn}-based data encoding, and a trainable~\ac{tn}-inspired~\ac{vqc} - we enable end-to-end training using a single classical optimiser, eliminating the need for pre-training circuit representations for each data point.

We could solve the binary classification task perfectly, and achieved promising results for the time series multivariate and univariate regression tasks. Although, the optimal choice of hyperparameters remains a challenge.

Future research directions include conducting an in-depth ablation study to explore the relationship between encoding parameters, such as bond dimension and number of disentangling layers, and the expressivity of the circuit. The classification task results suggest that entanglement/correlations in the compressed encoding data may not be essential for the classifier. However, it is unclear what portion of the Hilbert space the classifier/regressor accesses, potentially leaving room for improvement with encodings that offer more entanglement - a unique property of quantum computing. To clarify this point, future work will focus on analysing the~\ac{tn} representation of the processing layers, including examining the bond dimension of the~\ac{tn} representation of the quantum circuit, which provides insights into the amount of correlations present in the system.

Furthermore, our findings highlight the need for introducing more non-linearities into the hybrid model. This is a natural problem that occurs in quantum mechanics, being, at its core, a linear theory. To address this, we propose exploring techniques that interact with a classical environment, such as encoding, data re-uploading~\cite{PerezSalinas2020datareuploading}, natural noise models, or analogue mode operations present in~\ac{nisq} devices. These approaches have the potential to add sufficient non-linearities to the hybrid~\ac{qml} pipeline, leading to improved performance on aeroelastic problems. 

Overall, this study provides a foundation for future research in hybrid quantum tensor network-based algorithms for aeroelastic problems. By addressing the challenges and limitations identified in this work, we can unlock the full potential of these algorithms and explore their applications in more complex and realistic scenarios.

\backmatter

\bmhead{Acknowledgements}

We would like to express our sincere gratitude to Steffen Leger, Gustav J{\"a}ger, Andrew Barlow, and Krzysztof Bieniasz for their valuable contributions.

\bmhead{Author contribution}

Conceptualisation: M.L.H., D.Q., H-M.R; Methodology: M.L.H., P.A., H-M.R; Formal analysis and investigation: M.L.H., D.Q., H-M.R; Writing - original draft preparation: M.L.H.; Writing - review and editing: M.L.H., P.A., D.Q., F.S., H-M.R; Funding acquisition: D.Q., H-M.R; Data curation: D.Q.; Supervision: F.S., H-M.R.

\section*{Declarations}

\bmhead{Data availability}

Data sets utilised during the current study are available from the corresponding author on reasonable request. The simulations are available from the DLR but restrictions apply to the availability of these and are therefore not publicly available.

\bmhead{Conflict of interest} 

The authors declare no competing interests.

\bmhead{Funding}

M. Lautaro Hickmann received funding from the DLR Quantum Fellowship Program.

\bibliography{sn-bibliography}

\end{document}